\documentclass[epj]{svjour}

\usepackage{graphicx}
\usepackage{fancyhdr}
\def\makeheadbox{{
 }}
\def\mytitle{My title} 
\def\myauthors{My name}  
\def\mytype{My type of session}
\def\mysession{My session}

\def\mytitle{Lifetime Difference and CP Asymmetry in the $\mathbf{B_s \to J/\psi\phi}$ decay} 
\def\myauthors{Thomas Kuhr}    
\def\mytype{Contributed Talk}    
\def\mysession{Flavor Physics}

\begin{document}
\title{Lifetime Difference and CP Asymmetry in the $B_s \to J/\psi\phi$ decay}
\author{Thomas Kuhr\inst{1}
\thanks{\emph{Email:} Thomas.Kuhr@ekp.uni-karlsruhe.de}%
\ on behalf of the CDF Collaboration
}                     
%
%
\institute{Institut f\"ur Experimentelle Kernphysik, Universit\"at Karlsruhe, Wolfgang-Gaede-Str. 1, 76131 Karlsruhe, Germany}
%
\date{}
\abstract{
The $B_s$ meson is an interesting particle to study because a sizable mixing induced $CP$ violation
in the $B_s-\bar{B}_s$ system would be an indication for physics beyond the Standard Model.
In this paper we present a measurement of the lifetime difference $\Delta\Gamma$ between the $B_s$ mass 
eigenstates and the $CP$ violating phase in the decay $B_s \to J/\psi\phi$.
In 1.7 fb$^{-1}$ of data collected with the CDF II detector at the Tevatron $p\bar{p}$ 
collider we measure $\Delta\Gamma=0.076^{+0.059}_{-0.063}$ (stat.) $\pm 0.006$ (syst.) ps$^{-1}$,
well consistent with the Standard Model prediction, and a mean $B_s$ lifetime of
$c\tau_s=456 \pm 13$ (stat.) $\pm 7$ (syst.) $\mu$m.
We find no evidence for $CP$ violation \cite{CDFphis,CDFnote8950}.
\PACS{
      {13.20.He}{Decays of bottom mesons}   \and
      {14.40.Nd}{Bottom mesons}
     } 
} 
\maketitle
%
\section{Introduction}
\label{intro}
In the $B_s$-$\bar{B}_s$ meson system the flavor eigenstates are not the same as the mass eigenstates.
The mass difference between the heavy and light mass eigenstate, $B_{sH}$ and $B_{sL}$, determines the frequency 
of the oscillation of the $B_s$ mesons.
Two other quantities which affect the time evolution of $B_s$ mesons are the decay rates $\Gamma_H$ and 
$\Gamma_L$ of the two mass eigenstates.
The difference $\Delta\Gamma = \Gamma_L - \Gamma_H$ was measured first by CDF\cite{Acosta:2004gt} 
and recently with higher precision by D\O\cite{Abazov:2007tx}.

If the difference $\Delta\Gamma$ is larger than a few percent of the mean decay rate 
$\Gamma = (\Gamma_L+\Gamma_H)/2$ a time dependent analysis of $B_s$ decays without flavor tagging becomes 
sensitive to a further quantity, the $CP$ violating phase $\phi_s$.
This phase describes the mixing induced $CP$ violation and is related to the angle $\beta_s$ in the nearly 
degenerated unitarity triangle obtained from the multiplication of the second and third column of the CKM matrix.
The Standard Model expectation value for $\phi_s$ is very small\cite{Lenz:2006hd}.
Therefore a measurement of the phase which deviates significantly from zero would indicate new physics.

To determine $\Delta\Gamma$ the lifetime distribution of $B_s$ decays is measured.
Because it is very challenging to distinguish the two components of the lifetime distribution additional 
information is needed to separate the light and heavy mass eigenstates.
Therefore we exploit the fact that the two mass eigenstates are related to the $CP$ eigenstates.
In case of no $CP$ violation ($\phi_s=0$) $B_{sH}$ is $CP$ odd and $B_{sL}$ is $CP$ even.

A decay mode that allows to measure both lifetimes is $B_s \to J/\psi \phi$ with $J/\psi \to \mu^+ \mu^-$ 
and $\phi \to K^+ K^-$ which is a composition of $CP$ even and odd states.
Because the $B_s$ is a pseudo scalar and $J/\psi$ and $\phi$ are vector mesons, the orbital angular momentum
between the two decay products can have the values 0, 1 or 2.
S- and D-wave decays are $CP$ even, P-wave decays are $CP$ odd.
Consequently, the two $CP$ eigenstates can be separated by their different angular distributions of the decay products.

\begin{figure}
\begin{center}
\includegraphics[width=0.5\textwidth]{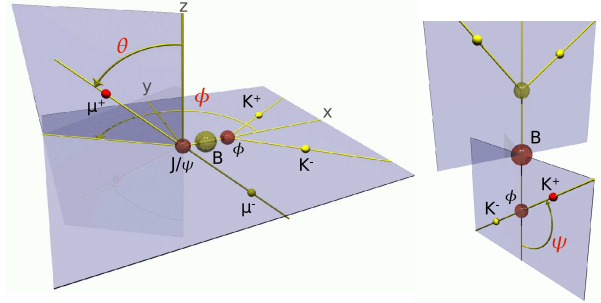}
\end{center}
\caption{Definition of transversity angles $\theta$, $\phi$ and $\psi$.}
\label{fig:transversity}
\end{figure}

The angles $\vec{\omega}=(\cos\theta,\phi,\cos\psi)$ used in this analysis are defined in the transversity 
basis illustrated in Figure \ref{fig:transversity}.
$\theta$ and $\phi$ are the polar and azimuthal angle of the $\mu^+$ in the rest frame of the $J/\psi$ 
where the $x$-axis is defined by the direction of the $B_s$ and the $xy$-plane by the $\phi \to K^+ K^-$ 
decay plane.
$\psi$ is the helicity angle of the $K^+$ in the $\phi$ rest frame with respect to the negative $B_s$ flight 
direction.

\section{Data Sample and Selection}
\label{sec:data}
The analyzed data sample with an integrated luminosity of 1.7 fb$^{-1}$ was collected by the
CDF II detector at the Tevatron which collides $p\bar{p}$ at a centre of mass energy of 1.96 TeV.
The detector components essential for this analysis are the silicon vertex detector for a precise
lifetime measurement, the central drift chamber for good momentum and mass resolution and
the central muon chambers for the identification and selection of muons.
In addition to the energy loss in the tracker the time of flight detector is used for particle
identification.
The $B_s \to J/\psi \phi$ events are triggered by a pair of two oppositely charged tracks matched
to signals in the muon chambers and having an invariant mass close to the $J/\psi$ mass.

\begin{figure}
\begin{center}
\includegraphics[width=0.45\textwidth]{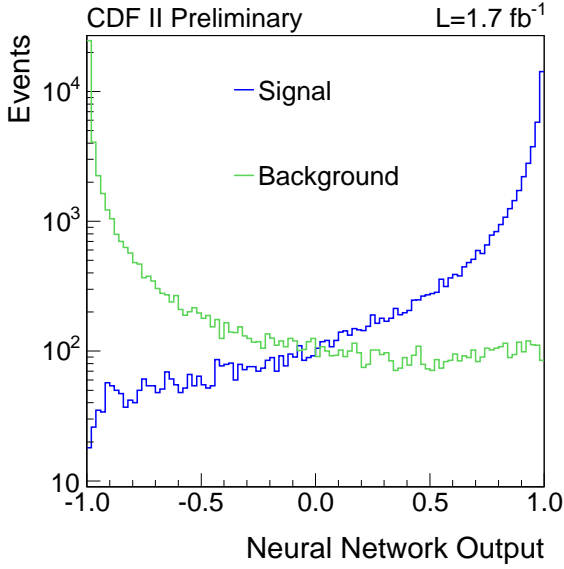}
\end{center}
\caption{Network output for $B_s$ signal and background events.}
\label{fig:NN}
\end{figure}

The $J/\psi$ candidates are combined with $\phi$ candidates built from pairs of oppositely charged 
tracks assumed to be kaons.
The $B_s$ candidate is obtained in a common vertex fit.
After applying basic kinematic cuts a neural network is used to improve the selection.
The network is trained on simulated $B_s$ signal events and background events from $B_s$ mass
sideband data.
Kinematic, particle identification and vertex fit quality variables are used as input to the network.
Figure \ref{fig:NN} illustrates the good separation of signal and combinatorial background events.

With a cut on the network output that optimizes the signal significance 2500 $B_s \to J/\psi\phi$ 
decays are selected.

\section{Mass, Lifetime and Angle Fit}
\label{sec:fit}
The mean $B_s$ lifetime $c\tau_s=c/\Gamma$, the lifetime difference $\Delta\Gamma$ and the decay
amplitudes $A_0$, $A_{||}$, and $A_\perp$ of the three angular components with relative phases
$\delta_{\perp}$ and $\delta_{||}$ are extracted in a 5-dimensional unbinned maximum likelihood fit 
in mass, lifetime and angular space.
Empirical models are used for the background distribution.
The $B_s$ mass signal is described by a sum of two Gaussians.
The lifetime and angle distribution is given by
\begin{eqnarray}
\frac{d^4P(\vec{\omega},t)}{d\vec{\omega}dt} &\propto& 
 |A_0|^2f_1(\vec{\omega}){\cal T}_+ + |A_{||}|^2f_2(\vec{\omega}){\cal T}_+\nonumber \\
 &+& |A_{\perp}|^2f_3(\vec{\omega}){\cal T}_- + |A_0||A_{||}|f_5(\vec{\omega}) \cos(\delta_{||}){\cal T}_+\nonumber \\
 &+& |A_{||}||A_\perp|f_4(\vec{\omega}) \cos(\delta_{\perp}-\delta_{||}) \nonumber \\
 & & \sin\phi_s (e^{-\Gamma_Ht}-e^{-\Gamma_Lt})/2 \nonumber \\
 &+& |A_{0}||A_\perp|f_6(\vec{\omega}) \cos(\delta_{\perp}) \nonumber \\
 & & \sin\phi_s (e^{-\Gamma_Ht}-e^{-\Gamma_Lt})/2
\label{eq:timedepangles}
\end{eqnarray}
with
\begin{eqnarray*}
{\cal T}_\pm &=& ((1\pm\cos\phi_s)e^{-\Gamma_Lt}+(1\mp\cos\phi_s)e^{-\Gamma_Ht})/2 \\
 f_1(\vec{\omega})&=& \frac{9}{32\pi} 2\cos^2\psi(1-\sin^2\theta \cos^2\phi)\\
 f_2(\vec{\omega})&=& \frac{9}{32\pi} \sin^2\psi(1-\sin^2\theta \sin^2\phi)\\
 f_3(\vec{\omega})&=& \frac{9}{32\pi} \sin^2\psi \sin^2\theta\\
 f_4(\vec{\omega})&=& -\frac{9}{32\pi} \sin^2\psi \sin2\theta \sin\phi\\
 f_5(\vec{\omega})&=& \frac{9}{32\pi} \frac{1}{\sqrt{2}}\sin2\psi \sin^2\theta \sin2\phi\\
 f_6(\vec{\omega})&=& \frac{9}{32\pi} \frac{1}{\sqrt{2}}\sin2\psi \sin2\theta \cos\phi.
\end{eqnarray*}
Note that this distribution is invariant under the transformations
\begin{eqnarray}
\phi_s \rightarrow -\phi_s, & & \delta_\perp \rightarrow \delta_\perp + \pi \quad \mbox{and} \nonumber \\
\Delta\Gamma \rightarrow -\Delta\Gamma, & & \phi_s \rightarrow \phi_s + \pi
\label{eq:ambiguity}
\end{eqnarray}
Because of this four fold ambiguity this measurement is insensitive to the sign of both,
$\phi_s$ and $\Delta\Gamma$.

\begin{figure}
\begin{center}
\includegraphics[width=0.42\textwidth]{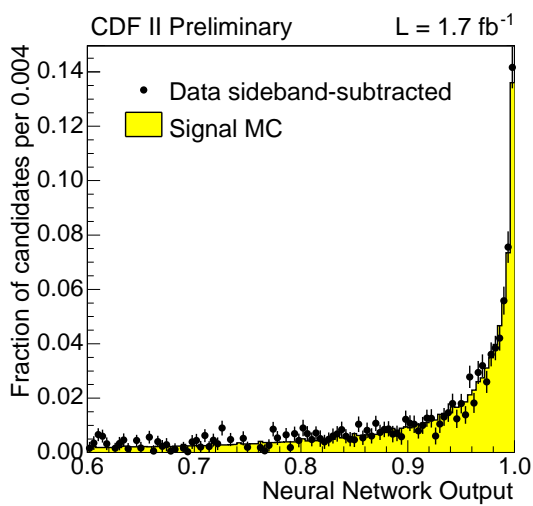}
\end{center}
\caption{Distribution of the selection network output for sideband subtracted data and simulation.}
\label{fig:dataMC}
\end{figure}

The finite lifetime resolution and differences of it between signal and background are included 
in the fit model.
The angle dependent acceptance is taken into account by an acceptance function obtained from
simulated events.
The good description of data by the simulation is exemplarily shown for the selection network output
in Figure \ref{fig:dataMC}.

\begin{figure}
\begin{center}
\includegraphics[width=0.45\textwidth]{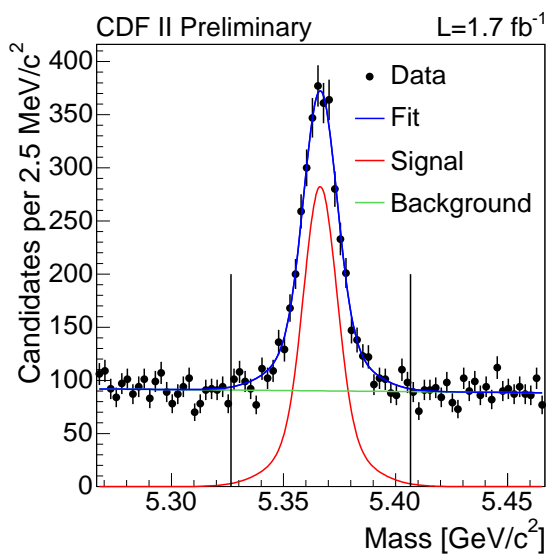}
\includegraphics[width=0.45\textwidth]{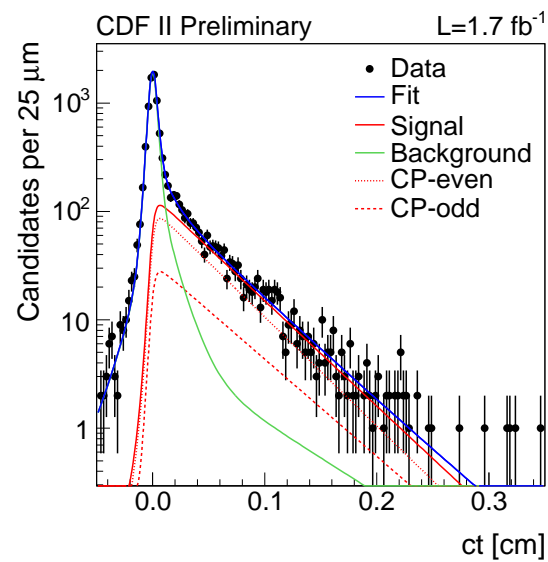}
\includegraphics[width=0.45\textwidth]{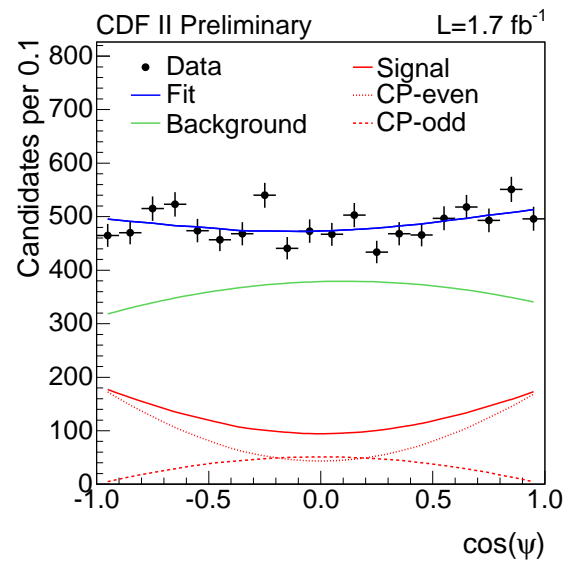}
\end{center}
\caption{Fit projections for mass, lifetime and angle $\cos\psi$.}
\label{fig:fit}
\end{figure}

The fit projections for mass, lifetime and the angle $\cos\psi$ are shown in Figure \ref{fig:fit}.

\section{Result Assuming no CP Violation}
\label{sec:phiS0}
We first consider the case of no $CP$ violation ($\phi_s=0$).
This simplifies the fit model because the last two terms in equation (\ref{eq:timedepangles})
and the fit parameter $\delta_\perp$ vanish.

The result can be affected by several systematic uncertainties which are evaluated using pseudo experiments.
The investigated effects are the influence of the angular background, the signal mass and
the lifetime resolution model, the contamination of misreconstructed $B^0 \to J/\psi K^*$ decays,
the acceptance function and the silicon detector alignment.
The largest systematic uncertainty for $\Delta\Gamma$ is the $B^0$ cross feed and for $c\tau_s$
the lifetime resolution model and the alignment.

\begin{figure}
\begin{center}
\includegraphics[width=0.42\textwidth]{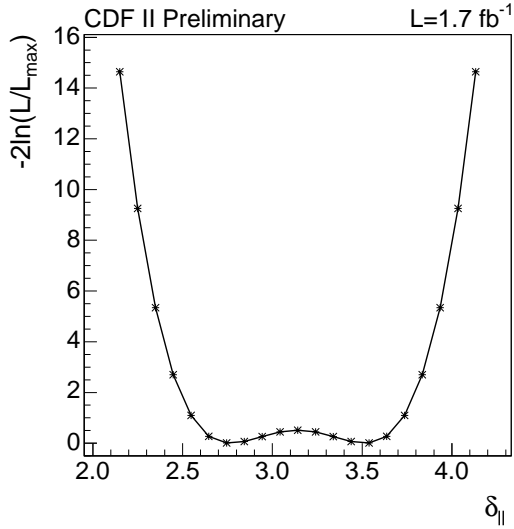}
\end{center}
\caption{Likelihood scan for the strong phase $\delta_{||}$.}
\label{fig:lh_deltaPar}
\end{figure}

Setting $\phi_s=0$ in the fit we obtain
\begin{eqnarray*}
	c\tau_s 		&=& 456 \pm 13 	 \pm 7\mbox{ $\mu$m} \\ 
	\Delta\Gamma 		&=& 0.076 ^{+0.059}_{-0.063} \pm 0.006 \mbox{ ps$^{-1}$} \\ 
	|A_0|^2 		&=& 0.530 \pm 0.021 \pm 0.007 \\ 
	|A_{||}|^2 		&=& 0.230 \pm 0.027 \pm 0.009
\end{eqnarray*}
The first is the statistical and the second one the systematic uncertainty.
Since the likelihood scan for the strong phase $\delta_{||}$, shown in Figure \ref{fig:lh_deltaPar},
has a non-parabolic shape due to a symmetry at $\delta_{||}=\pi$
we do not quote a point estimate for this quantity.

\section{Fit with Floating CP Violating Phase $\mathbf{\phi_s}$}
\label{sec:bias}
\begin{figure}
\begin{center}
\includegraphics[width=0.5\textwidth]{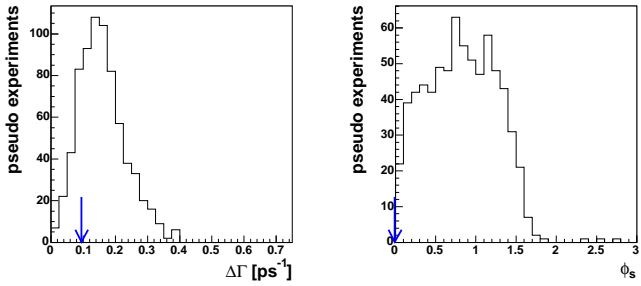}
\end{center}
\caption{Distribution of fitted $\Delta\Gamma$ (left) and $\phi_s$ (right) in pseudo experiments.
The blue arrow indicates the input value.}
\label{fig:bias}
\end{figure}

Since a maximum likelihood fit is only guaranteed to be unbiased in case of unlimited statistics
we studied the fit in pseudo experiments.
In case of free $\phi_s$ parameter we observe that for low input values of $\Delta\Gamma$ or $\phi_s$ 
there is a bias towards higher values.
This is illustrated in Figure \ref{fig:bias}.

The bias can be understood by looking at equation (\ref{eq:timedepangles}).
If $\phi_s$ approaches zero the last two terms vanish and $\delta_\perp$ becomes undetermined.
This means that the fit can not improve the description of the data any more by varying $\phi_s$.
It effectively lost a degree of freedom.
The situation is similar when $\Delta\Gamma$ gets zero.
Then $\phi_s$ and $\delta_\perp$ are undetermined.

Because of the biased fit result we do not quote a point estimate for $\Delta\Gamma$ and $\phi_s$,
but construct a confidence region following the procedure suggested by Feldman and Cousins\cite{Feldman:1997qc}.

For each pair on a grid of assumed true values of $\Delta\Gamma$ and $\phi_s$ we calculate a $p$-value,
which quantifies the probability to get the fit result observed in data.
To determine the $p$-value we use the likelihood ratio
\begin{eqnarray}
R(\Delta\Gamma, \phi_s)&=&\log \frac{{\cal L}(\hat{\Delta\Gamma}, \hat{\phi_s}, \hat{\theta})}{{\cal L}(\Delta\Gamma, \phi_s, \hat{\theta}^\prime)}
\end{eqnarray}
where $\theta$ are the nuisance parameters and the hat indicates the parameter values which maximize the
likelihood $\cal{L}$.
The $R$ distribution for assumed true values of $\Delta\Gamma$ and $\phi_s$ is obtained from pseudo
experiments.
The pseudo experiment input values of the nuisance parameters are taken from a fit to data,
a procedure known as plug-in method.

The $p$-value is the fraction of pseudo experiments with an $R$ value higher than the one in data.
The 90\% (95\%) confidence region is then defined by the $\Delta\Gamma$-$\phi_s$ points with a $p$-value
above 10\% (5\%).

\begin{figure}
\begin{center}
\includegraphics[width=0.5\textwidth]{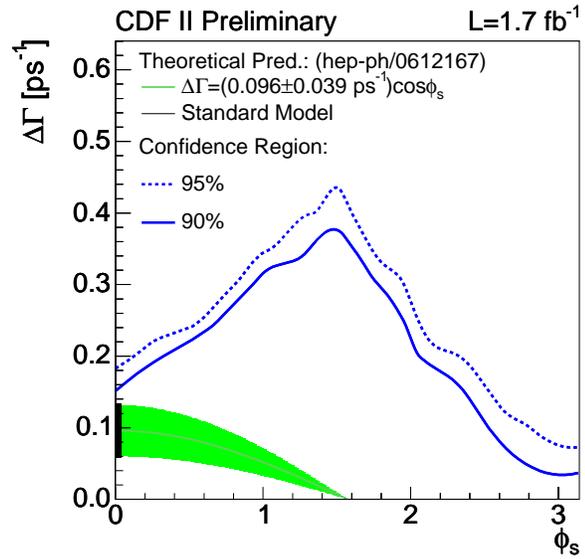}
\end{center}
\caption{Confidence region in the $\Delta\Gamma$-$\phi_s$ plane.
Only one out of four possible solutions is shown.
The others can be obtained by the transformations $\phi_s \to -\phi_s$
and $\Delta\Gamma \to -\Delta\Gamma$, $\phi_s \to \phi_s + \pi$}
\label{fig:confidence}
\end{figure}

The result is presented in Figure \ref{fig:confidence}.
Note that only the first quadrant is shown.
The other three quadrants can be obtained via the transformations given in equation (\ref{eq:ambiguity}).
The $p$-value for $\Delta\Gamma=0.1$ ps$^{-1}$ and $\phi_s=0$, which are approximately the Standard
Model predictions, is 22\%.

\section{Conclusions}
\label{sec:conclusions}
Using 2500 $B_s \to J/\psi\phi$ decays selected by a neural network in a data sample of
1.7 fb$^{-1}$ CDF has performed a mass-lifetime-angle fit to measure the lifetime
difference $\Delta\Gamma$ between the $B_s$ mass eigenstates.
The value obtained under the assumption of no $CP$ violation is consistent with the Standard Model
expectation\cite{Lenz:2006hd} and previous measurements\cite{Acosta:2004gt,Abazov:2007tx}.
The extracted mean $B_s$ lifetime is currently the most precise measurement and agrees well
with the world average $B^0$ lifetime as predicted by theory.

If we allow for $CP$ violation in the fit model we observe a bias away from low $\Delta\Gamma$
and $\phi_s$ values.
It is understood by the structure of the likelihood function and limited statistics.
Instead of a point estimate a confidence region is determined in a frequentist way.
The result is compatible with the Standard Model and can not rule out any minimal flavor violating new physics 
scenario which changes the phase $\phi_s$, but does not significantly affect 
$b \to c\bar{c}s$ tree level dominated processes.

%

\begin{thebibliography}{999}
\bibitem{CDFphis}
  The $CP$ violating phase measurement was presented by CDF briefly after the SUSY07 conference 
  and is therefore included in this article.
\end{thebibliography}
%

\end{document}